\documentclass[conference]{IEEEtran}
\usepackage{amsmath, amssymb, amsthm}
\usepackage{graphicx}
\usepackage{cite}
\usepackage{color}
\usepackage{balance}
\usepackage{booktabs}
\usepackage{multirow}
\usepackage {tikz}
\usetikzlibrary {positioning}
\usetikzlibrary{arrows.meta}

\newcommand{\figw}{0.92\linewidth}
\begin{document}

\title{Age of Information in Queueing Systems with Merging Server Streams
}

\author{
\IEEEauthorblockN{Lucrezia Rossi, Leonardo Badia}
\IEEEauthorblockA{{Dept.\ Information Engineering} \\
University of Padova, Padua, Italy \\
\{rossilucrezia, badia\} @dei.unipd.it}
\and
\IEEEauthorblockN{Andrea Munari}
\IEEEauthorblockA{German Aerospace Center (DLR)\\ We{\ss}ling, Germany \\
andrea.munari@dlr.de}
}

\maketitle

\begin{abstract}
We present a novel analytical framework for evaluating the age of information (AoI) in multi-path queueing topologies where output streams from multiple servers converge into a single pipeline. This theoretical scenario finds immediate application in ultra-reliable low-latency networks and edge computing architectures, where multi-path routing and stream merging can be vital strategies for maintaining fresh status updates. Specifically, we investigate two setups: a split-merge network and a duplicate-merge network. The split-merge case has been addressed in prior work, but we show that existing studies are incorrect, and our approach provides instead an exact analytical formulation. Moreover, building on similar reasoning, we provide the first formal analysis of the previously unexplored duplicate-merge scenario. Finally, we show how to leverage our theoretical results to solve key system-level optimization problems, deriving the optimal randomized routing probabilities and traffic injection rates that minimize average AoI.
\end{abstract}
\section{Introduction}
\label{sec:intro}

The advent of ultra-reliable low-latency communications in next-generation networks is driving the deployment of applications that rely on real-time information about the surrounding environment \cite{popovski2022perspective}. In such systems, the freshness of information is a key performance indicator and is commonly quantified through metrics such as age of information (AoI) \cite{yates2021age}. The precise mathematical formulation of AoI has revitalized fundamental research at the intersection of queueing theory, optimization, and stochastic analysis \cite{fidler2023statistical,reddy2025age}.

At the same time, multiaccess edge computing (MEC) and other distributed processing architectures require computational tasks to be distributed across multiple heterogeneous servers \cite{jin2022review}. The resulting parallel processing hierarchy often requires data to be split or duplicated to avoid congestion bottlenecks, with the outputs subsequently merged to produce the final result. While this approach improves scalability and resource utilization, it also substantially increases the analytical complexity of the system, making the evaluation of AoI-like metrics in converging architectures difficult \cite{tang2023age}.

\begin{figure}[tbp]
    \centering
    {%
    \includegraphics[width=0.48\columnwidth]{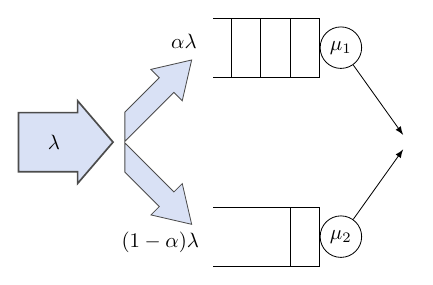}%
    }
    {%
    \includegraphics[width=0.48\columnwidth]{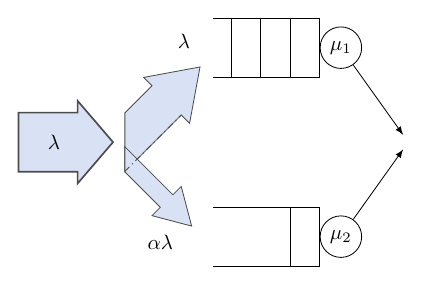}
    }
    \caption{Considered scenarios. Left: split-merge. Right: duplicate-merge.}
    \label{fig:code}
\end{figure}

Consequently, quantifying the freshness of information delivery in merging systems is not only a non-trivial mathematical problem, but also a fundamental design issue for future communication networks. To address this challenge, we present a novel analytical framework to evaluate the Age of Information in multi-path queueing topologies where parallel server outputs converge into a single processing line. Specifically, we focus on two different fork-split paradigms: a \emph{split-merge} architecture and a \emph{duplicate-merge} architecture involving M/M/1 queues, represented in Fig.\ \ref{fig:code}.

The case of split-merge implies that two M/M/1 queues are employed to serve status updates coming from a single source. The traffic is randomly divided, each packet being routed to server $1$ with probability $\alpha$ or server $2$ with probability $1{-}\alpha$. This directly relates to the stateless random-alpha routing adopted in MEC \cite{mancuso2022stateful,cohen2023dynamic}.
Although the AoI for this topology has been studied in the prior literature \cite{bhati2021age}, existing analytical techniques suffer from high structural complexity, which forced the introduction of some hidden assumptions. We will show that this causes the existing models to become incorrect at high loads. In contrast, our methodology yields an exact closed-form analytical formulation.

Building upon this framework, we also provide the analysis of the duplicate-merge scenario, where traffic is copied on a secondary path with probability $\alpha$, rather than split in an alternative fashion. As a result, the first queue receives $100\%$ of the traffic, while the second queue only a fraction $\alpha$ of it, and its packets are always present in the other queue. To our knowledge, this case has never been attempted before, and we provide its first exact formal AoI analysis. 

Beyond performance evaluation, we demonstrate the practical utility of our theoretical results by showing how they can be used for system-level optimization problems, such as deriving the optimal randomized routing probabilities and traffic injection rates required to minimize the average AoI.

The remainder of this paper is organized as follows.
Section \ref{sec:relw} discusses the related literature. Section \ref{sec:split-model} analyzes the split-merge network. Section \ref{sec:analysis_duplicate} presents the analysis of the duplicate-merge topology. We present numerical results in Section \ref{sec:result} and conclude in Section \ref{sec:conclusion}.

\section{Related Work}
\label{sec:relw}
AoI, introduced in \cite{kaul2012real}, has emerged as a fundamental metric for quantifying information freshness in real-time monitoring and status update systems. Initial investigations of multi-path networks demonstrated that transmission path diversity can substantially mitigate status staleness by providing alternative parallel routes \cite{kam2015effect}. To analyze complex queueing dynamics in such multi-source scenarios, Yates and Kaul developed a general analytical framework based on stochastic hybrid systems (SHS) \cite{yates2018age}, which became the cornerstone for computing AoI in networks with multiple updating streams.

These investigations about the interaction between packet generation policies and service capacities in multi-source environments were also revisited by the same authors for multi-source $M/M/1$ queues to characterize optimal update rates under preemptive and non-preemptive disciplines in \cite{kaul2020timely}, while Moltafet \textit{et al.} provided a more precise AoI evaluation across various multi-source queueing models in \cite{moltafet2020age} and also extended the analysis to a wide variety of queueing systems in \cite{moltafet2021average}, but always considering multiple \emph{independent} sources, each generating different traffic.
The influential summary on the main findings for AoI-related metrics reported in \cite{yates2021age} discusses many results, but most of them relate to the case of independent sources and multiple servers in the same facility.

Subsequent efforts expanded these frameworks to structured queueing topologies where multiple sources are not independent but rather pertain to similar observations. 
A first line of research involves the study of \emph{correlated} information, often motivated by the observation that multiple monitoring sources often measure overlapping portions of the surrounding environment and/or similar metrics. In this spirit, Tripathi and Modiano \cite{tripathi2022optimizing} as well as Tong \emph{et al.} \cite{tong2022age} proposed optimization frameworks for source scheduling, while Zancanaro \emph{et al.} \cite{zancanaro2023modeling} studied a stochastic model similar to an M/M/1 queue. A more recent extension by Liang and Zhou \cite{liang2025analysis} included a random access in the analysis, and Fidler \emph{et al.} connected the correlation to geometry aspects in \cite{fidler20252d}, still making explicit reference to classic queueing theory. All of these contributions, even when directly addressing queues, present, however, fundamentally different models in that the system always consists of multiple independent sources. The correlation of content just considers some packets as informative for other sources as well. However, preemption issues are not considered and the presence of additional flows is always advantageous.

For the specific case where the interdependency of content stems instead from an initial separation among multiple paths and a subsequent merge along the way, Chiariotti \textit{et al.} analyzed the impact of non-preemptive tributary traffic converging onto a main transmission line, highlighting the queueing delays induced by merging streams \cite{chiariotti2022age}. In \cite{tang2023age}, Tang \emph{et al.} studied a similar splitting policy for MEC, but where the probabilistic split (the third policy examined in the paper) revolves around a queue tandem representing different stages of the processing of subtasks, instead of a split-merge lane. For stream convergence, the performance of parallel transmission channels was examined by Fidler \textit{et al.}, who established statistical AoI limits to quantify when independent channel diversity yields tangible freshness gains \cite{fidler2023statistical}. 

In a similar vein, Bhati \textit{et al.} derived AoI expressions for queueing systems with parallel heterogeneous servers \cite{bhati2021age}.
This last paper presents an analysis similar to ours; however, it subtly introduces approximations that render it inexact, leading to discrepancies in the AoI values. In contrast, our analysis is exact and, as we will show, resolves these issues.

More recently, Chen \textit{et al.} analyzed status update timeliness when parallel updates are sent across heterogeneous update devices \cite{chen2024timeliness}, and Reddy and Venkatesh explored proactive management strategies to discard obsolete status packets in multi-hop multi-source queues \cite{reddy2025age}. 
The findings above have had repercussions on cross-disciplinary issues such as stochastic control methods for $M/M/k$ multi-server queues. In this case, the optimization framework proposed by Lefebvre and Yaghoubi \cite{lefebvre2026optimal}, provides queue-management techniques for dynamic server activation tuning the value of $k$. 

Despite these extensive contributions, the existing literature does not have an exact analysis of AoI in either of these topologies. Thus, our work provides a tractable, exact analytical framework of both split-merge and duplicate-merge, correcting the existing models of the literature for the former, and solving the latter, which is an entirely open problem.

\section{Split-Merge: Analytical Derivation}
\label{sec:split-model}
 
We consider a split-merge (also sometimes called fork-split) status update system \cite{bhati2021age} in which a single source generates
status-update packets according to a Poisson process of rate $\lambda$. Upon
generation, each packet is independently routed to one of two parallel
first-come-first-served (FCFS), non-preemptive $M/M/1$ queues: queue~1, with
service rate $\mu_1$, is selected with probability $\alpha$, while queue~2, with
service rate $\mu_2$, is selected with the complementary probability $1-\alpha$.
Both sub-streams are therefore Poisson, with rates $\lambda_1=\alpha\lambda$ and
$\lambda_2=(1-\alpha)\lambda$ respectively, and each queue is stable if
$\lambda_i<\mu_i$, $i=1,2$.
 
Since both queues operate under the FCFS discipline, 
a packet can never be
overtaken by another packet queued behind it in that same queue. However,
because the two queues are physically separate and are read independently by
the monitor, a queue~$i$ packet can be overtaken, in terms of
delivery order, by a packet that was generated later but was routed to the
competing queue $j\neq i$ and happens to complete service first. When this
occurs, the overtaken packet no longer carries the most recent information
available to the monitor and, upon its (out-of-order) delivery, does not
reduce AoI. Only packets that are \emph{not}
overtaken, which we term \emph{informative}, actually reset AoI at
the monitor.
 
The goal of this work is to characterize the long-term average AoI of this
split-merge system. Following the standard graphical (area-based) decomposition
of the age process~\cite{yates2021age}, the average AoI can be written in terms
of two quantities associated with the sequence of informative departures: the
system time $S$ of an informative packet (i.e., the total time it spends in the
system, from generation to delivery) and the inter-departure time $T$ between
two consecutive informative departures, which is precisely the interval over
which the AoI grows linearly before being reset. The general expression for the
time-average AoI is
\begin{equation}
\Delta_{\mathrm{avg}} = \frac{\mathbb{E}[ST] + 
\mathbb{E}[T^2]/2}{\mathbb{E}[T]},
\label{eq:aoi-general}
\end{equation}
where the expectation is taken with respect to the stationary distribution of
the informative-departure process.
 
Two features distinguish this system from a classical single-queue analysis.
First, not every generated packet is informative, since packets that are
overtaken by the competing queue must be discarded from the age-relevant point
of view; this discarding mechanism is not statistically independent across
consecutive packets, so the aggregate process of informative departures is not
Markovian. Second, and as a direct consequence, $T$
must be understood in~\eqref{eq:aoi-general} as the inter-\emph{departure} time between informative
packets, and not as the inter{arrival} time of the underlying Poisson
source \cite{chiariotti2022age}. In our analysis, they do not coincide, because the filtering induced by the competing queue breaks
the memoryless renewal structure of the raw arrival process.

 
The starting point of the analysis is the probability that a packet is
overtaken by the competing queue, which can be derived via direct combinatorial conditioning on the queue-length process.
Consider a tagged packet routed to queue~$i$. It is overtaken if and only if
the competing queue $j$ generates a new packet and completes its service
before the tagged packet's own service is completed. Denote by $X_j$ the time
until the next arrival in queue $j$ and by $S_j$ the system time of that
competing packet. The relevant quantity is the compound random variable
$X_j+S_j$, which we refer to as the \emph{competing completion time}, since it
is the time required for queue $j$ to produce a new, potentially overtaking
departure. Its Laplace--Stieltjes transform (LST) is
\begin{IEEEeqnarray}{l}
x_f(s,\lambda_j,\mu_j) = \frac{\mu_j{-}\lambda_j}{(\mu_j{+}s)^2}
\!\left[\frac{\lambda_j^2}{\mu_j{+}s{-}\lambda_j} {+} \frac{\lambda_j(\mu_j{+}\lambda_j{+}s)}{s{+}\lambda_j}\right]\!.\!
\label{eq:xf} \qquad
\end{IEEEeqnarray}
Because the tagged packet's own system time $S_i$ is exponentially distributed
with rate $\nu_i=\mu_i-\lambda_i$, the overtaking probability follows directly
from the defining property of the Laplace transform, $\mathbb{E}[e^{-\nu Y}]=
\mathcal{L}\{Y\}(\nu)$ for $S_i\sim\mathrm{Exp}(\nu_i)$ independent of $Y$: the
probability that the exponential clock of the tagged packet outlasts the
competing completion time is obtained simply by evaluating~\eqref{eq:xf} at
$s=\nu_i$. Hence the overtaking probabilities of the two queues are
\begin{subequations}
\label{eq:U1U2}
\begin{align}
\nonumber
U_1 &= x_f(\mu_1-\lambda_1,\,\lambda_2,\,\mu_2), \\
\nonumber
U_2 &= x_f(\mu_2-\lambda_2,\,\lambda_1,\,\mu_1),
\end{align}
\end{subequations}
and the probability that a packet routed to queue $i$ is informative (i.e.,
not overtaken) is $1-U_i$.
 
Although the sequence of informative departures is not a renewal process,
since the thinning induced by the competing queue breaks the memoryless
structure of the underlying process, its \emph{mean} inter-departure time can
nonetheless be obtained without any knowledge of the full distribution of $T$,
through a simple flow-conservation (Little's law type) argument. For a
stationary and ergodic process, the long-run average time between consecutive
informative departures is the reciprocal of the long-run rate at which such
departures occur:
\begin{equation}
\mathbb{E}[T] = 1/{\lambda_{\mathrm{inf}}}.
\label{eq:ET}
\end{equation}
Since queue $i$ receives packets at rate $\lambda_i$ and a fraction $1-U_i$ of
them are informative, the informative-departure rate contributed by each
queue is
\begin{equation}
\lambda_{\mathrm{inf},1} = \lambda_1(1-U_1), \qquad
\lambda_{\mathrm{inf},2} = \lambda_2(1-U_2),
\end{equation}
and 
the aggregate informative-departure rate is
the sum of the two queue-level rates,
\begin{equation}
\lambda_{\mathrm{inf}} = \lambda_{\mathrm{inf},1}+\lambda_{\mathrm{inf},2}
= \lambda\bigl[\alpha(1-U_1)+(1-\alpha)(1-U_2)\bigr].
\label{eq:lambda-inf}
\end{equation}
This argument only requires knowledge of the \emph{rate} of informative
departures and is therefore insensitive to the fact that $T$ is not
exponentially, nor even memorylessly, distributed. The remaining moments,
$\mathbb{E}[ST]$ and $\mathbb{E}[T^2]$, instead require the full distribution
of $T$ and are derived next.
 
Unlike $\mathbb{E}[T]$, the cross moment $\mathbb{E}[ST]$ cannot be obtained
from a rate argument alone, because $S$ and $T$ are statistically dependent: a
longer system time $S$ for the tagged, informative packet leaves more time for
the competing queue to generate a new overtaking candidate, and it also
affects the residual state of the tagged packet's own queue at the moment of
departure. Both effects must be accounted for jointly.
 
We condition on the system time of the tagged informative packet, $S=s$. Given
$S=s$, the subsequent inter-departure time $T$ exceeds $t$ if and only if two
independent events both hold: (i) the competing queue $j$, already known not
to have overtaken the tagged packet by time $s$, continues not to produce a
new overtaking completion up to time $s+t$; and (ii) the tagged packet's own
queue $i$ does not itself produce another departure within the following $t$
units of time.
 
The first event is governed by the survival function of the competing
completion time,
\begin{equation}
R_j(u) := \Pr(X_j+S_j > u),
\end{equation}
whose Laplace transform follows from~\eqref{eq:xf} via
$\mathcal{L}\{R_j\}(s)=\bigl(1-x_f(s,\lambda_j,\mu_j)\bigr)/s$, and which admits
the closed form
\begin{equation}
\begin{aligned}
R_j(u) = {} & e^{-(\mu_j-\lambda_j)u} - \lambda_j u\, e^{-\mu_j u} \\
& - \frac{\mu_j}{\mu_j-\lambda_j}e^{-\mu_j u} + \frac{\mu_j}{\mu_j-\lambda_j}e^{-\lambda_j u}.
\end{aligned}
\label{eq:Rj}
\end{equation}
Conditioned on survival past $s$, the probability that queue $j$ remains
silent up to $s+t$ is the ratio $R_j(s+t)/R_j(s)$.
 
The second event depends on the state of queue $i$ left behind by the
departing tagged packet. If no arrival occurred in queue $i$ during $[0,s]$,
the queue is empty and the next departure requires a fresh arrival followed by
a service; otherwise, an arrival is already waiting and the next departure is
simply a residual, memoryless service. This yields the own-queue survival
function
\begin{IEEEeqnarray}{l}
R_i(t \mid s) = e^{-\lambda_i s}\,\frac{\mu_i e^{-\lambda_i t}{-}\lambda_i e^{-\mu_i t}}{\mu_i-\lambda_i} + \!\bigl(1{-}e^{-\lambda_i s}\bigr)e^{-\mu_i t}. \;\;\;\;\;\;
\label{eq:Ri}
\end{IEEEeqnarray}
In~\eqref{eq:Ri}, the first term, weighted by $e^{-\lambda_i s}$ (the
probability that no arrival occurred in queue $i$ during $[0,s]$), corresponds
to the case in which the queue is left empty, so that the next departure
requires a new arrival followed by a full service, yielding a hypoexponential
residual time; the second term, weighted by the complementary probability
$1-e^{-\lambda_i s}$, corresponds to the case in which an arrival is already
waiting, so that the next departure is simply a memoryless residual service of
rate $\mu_i$.
 
Since the two queues are independent, the two survival probabilities
multiply,
\begin{equation}
\Pr(T>t \mid S=s) = \frac{R_j(s+t)}{R_j(s)}\,R_i(t\mid s).
\end{equation}
Averaging this expression over the (informative) density of $S$,
$f_{S,i}(s) = g_i e^{-g_is}R_j(s)/P_{\mathrm{inf},i}$ with $g_i=\mu_i-\lambda_i$
and $P_{\mathrm{inf},i}=1-U_i$, the factor $R_j(s)$ cancels between numerator
and denominator, leaving
\begin{equation}
\Pr(T>t \mid i) = \frac{1}{P_{\mathrm{inf},i}}\int_0^\infty g_i e^{-g_i s}\,R_j(s+t)\,R_i(t\mid s)\, ds.
\label{eq:survT}
\end{equation}
 
To obtain $\mathbb{E}[ST\mid i]$ we invoke the law of total expectation,
$\mathbb{E}[ST\mid i] = \mathbb{E}\bigl[S\cdot\mathbb{E}[T\mid S,i]\bigr]$,
together with the standard survival-to-mean identity
$\mathbb{E}[T\mid S=s,i]=\int_0^\infty \Pr(T>t\mid S=s,i)\,dt$. Combining the
two and using~\eqref{eq:survT} gives
\begin{equation}
\mathbb{E}[ST\mid i] = \frac{1}{P_{\mathrm{inf},i}}\int_0^\infty\!\!\int_0^\infty
s\, g_i e^{-g_i s}\, R_j(s+t)\, R_i(t\mid s)\, ds\, dt.
\label{eq:EST}
\end{equation}
 
Both $R_j(\cdot)$ and $R_i(\cdot\mid s)$ are finite sums of terms of the form
$c\,u^k e^{-au}$ ($k\in\{0,1\}$), as seen in~\eqref{eq:Rj} and~\eqref{eq:Ri}.
Expanding $R_j(s+t)$ via the binomial theorem and multiplying out the product
in~\eqref{eq:EST} reduces the double integral to a finite sum -- fifteen terms
in total -- of elementary integrals of the form
\begin{equation}
\int_0^\infty s^p e^{-\alpha s}\, ds \cdot \int_0^\infty t^q e^{-\beta t}\, dt
= \frac{p!}{\alpha^{p+1}}\cdot\frac{q!}{\beta^{q+1}},
\label{eq:gamma-int}
\end{equation}
with $\alpha,\beta>0$ guaranteed by stability. Therefore,~\eqref{eq:gamma-int} yields $\mathbb{E}[ST\mid i]$ in closed form, without any numerical
integration.
 
Finally, the system-level moment is obtained by weighting the queue-level
results by the long-run share of informative departures
from each queue, $\pi_i = \lambda_i P_{\mathrm{inf},i}/(\lambda_1
P_{\mathrm{inf},1}+\lambda_2 P_{\mathrm{inf},2})$: 
\begin{equation}
\mathbb{E}[ST] = \pi_1\,\mathbb{E}[ST\mid 1] + \pi_2\,\mathbb{E}[ST\mid 2].
\label{eq:EST-final}
\end{equation}
 
The second moment of the inter-departure time follows the same construction as
$\mathbb{E}[ST]$, but does not require conditioning on $S$: the
survival function $\Pr(T>t\mid i)$ derived in~\eqref{eq:survT} already
incorporates the averaging over $S$ and can be used directly. We apply
the identity relating the second moment of a non-negative random
variable to its survival function \cite{chakraborti2019higher},
\begin{equation}
\mathbb{E}[T^2\mid i] = 2\int_0^\infty t\, \Pr(T>t\mid i)\, dt,
\end{equation}
which, substituting~\eqref{eq:survT}, gives
\begin{equation}
\mathbb{E}[T^2\mid i] = \frac{2}{P_{\mathrm{inf},i}}\int_0^\infty\!\!\int_0^\infty
t\, g_i e^{-g_i s}\, R_j(s+t)\, R_i(t\mid s)\, ds\, dt.
\label{eq:ET2-cond}
\end{equation}

Equation~\eqref{eq:ET2-cond} has the same integrand as~\eqref{eq:EST}, with the
extra multiplicative factor $t$ in place of $s$ and an overall factor of $2$
inherited from the survival-to-second-moment identity. The same
binomial-expansion argument applies unchanged: since $R_j(s+t)$ and $R_i(t\mid
s)$ are finite sums of terms $c\,u^k e^{-au}$, the double integral
in~\eqref{eq:ET2-cond} reduces to a finite sum of terms of the
form~\eqref{eq:gamma-int}, yielding $\mathbb{E}[T^2\mid i]$ in exact closed
form. The system-level second moment is obtained by the same throughput-share
weighting as in~\eqref{eq:EST-final},
\begin{equation}
\mathbb{E}[T^2] = \pi_1\,\mathbb{E}[T^2\mid 1] + \pi_2\,\mathbb{E}[T^2\mid 2].
\label{eq:ET2}
\end{equation}
 
Together, \eqref{eq:ET}, \eqref{eq:EST-final}, and~\eqref{eq:ET2} provide all
the ingredients required to evaluate the average AoI of~\eqref{eq:aoi-general} in closed form.

\section{Duplicate-Merge: Analytical Derivation}
\label{sec:analysis_duplicate}

We now turn to the case of duplicate-merge. Rather than being routed to either of
the two queues, every packet is now guaranteed to be delivered through queue~1 and
additionally \emph{duplicated} to queue~2 with probability $\alpha$. As in
the splitting model, packet is informative only if it is
not overtaken by a fresher packet from the competing queue. The
average AoI is still given by~\eqref{eq:aoi-general}. The two policies
differ, however, in a way that is consequential for the analysis: in the
splitting model the two queues are fed by independent Poisson sub-streams,
whereas here queue~2's arrivals are a sub-sample of queue~1's.
This coupling leaves the marginal behavior of each queue unchanged but makes
the queues' occupancies correlated, which in turn requires two
structurally different derivations: one for the overtaking probability and one for the joint law governing the case decomposition introduced below.

Since packets arrive according to a Poisson process of rate $\lambda$, and they are all routed
to queue~1 but independently, each packet is also duplicated to queue~2 with
probability $\alpha$, by Poisson thinning, the duplicated sub-stream is
itself Poisson of rate $\lambda':=\alpha\lambda$, and we write
$\nu_1:=\mu_1-\lambda$, $\nu_2:=\mu_2-\lambda'$, $\rho_2:=\lambda'/\mu_2$ for
the stability-guaranteed rates and traffic intensity of the two queues.
The departure process decomposes into three mutually exclusive
cases according to which copy of the record-setting packet determines the
next informative departure: Case~1, in which the record-setting packet was
not duplicated; and Cases~2 and~3, in which the packet was duplicated and the first to depart is the 
queue-1 copy or the queue-2 copy, respectively.

For each queue $j\in\{1,2\}$ we mimic the analysis of the split-merge scenario by considering that, after an informative packet enters the system at time $0$ and exits at time $Z$, the next departure time is found by considering two \emph{candidate times} ${C}_j = \max(X_j,S_j)+Y_j$, for $j=\{1,2\}$, with $Y_j\sim
\mathrm{Exp}(\mu_j)$ being the service time in queue $j$, $S_j$ the system time in queue $j$, and $X_j$ the arrival
epoch of the first subsequent packet
routed through queue $j$ after time $0$; the inter-departure time $T$ of informative packets is found from $\min({C}_1,{C}_2)-Z$. Indeed, if a packet is not overtaken, the minimum of ${C}_1$ and ${C}_2$ will be the exit time of the first subsequent informative packet.
We note that the analysis changed depending on the cases, as in Case 1 it must be that $Z=S_1$, as the non-duplicated packet can only exit through queue $1$, but also $S_1 < X_2 + S_2$ as in the split-merge case. In cases $2$ and $3$, instead, two duplicate packets enter at the same time (taken as time $0$). Of these, only one will be informative, and it will be the earliest of the two, so $Z = \min(S_1,S_2)$, as will be discussed next.

\subsection*{Case 1: non-duplicated record}

The tagged packet occupies queue~1 alone, with system time
$S\sim\mathrm{Exp}(\nu_1)$, and is overtaken if and only if
${C}_2$ completes before $S_1$. Since the tagged packet is not
duplicated, its departure epoch is by construction uncorrelated with
queue~2, so that queue~2's own arrival process, viewed from this reference
epoch, is exactly Poisson of rate $\lambda'$ -- statistically
indistinguishable, at the level of ${C}_2$ alone, from the
competing queue of the splitting model. The overtaking probability for
Case~1 therefore follows directly from the same competing-completion-time
transform used in Section~\ref{sec:split-model}, with the competing rate
$\lambda_j$ replaced by $\lambda'$ and evaluated at $s=\nu_1$:
\begin{equation}
	U_1 := x_f(\nu_1,\lambda',\mu_2),
	\qquad
	P_{\mathrm{inf},1} := 1-U_1,
	\label{eq:Pinf1-xf}
\end{equation}
with $x_f$ as defined in \eqref{eq:xf}, exactly as $U_i$ and $1-U_i$ were
introduced in~\eqref{eq:U1U2}. This is a genuine identity and
not an approximation, as the full construction of
$P_{\mathrm{inf},1}$ specialized to $w=0$ reproduces
\eqref{eq:Pinf1-xf} exactly.

As in the split-merge model, the mean inter-departure time of informative packets requires no knowledge
of the distribution of $T$ beyond its rate. Since a fraction
$P_{\mathrm{inf},1}$ of the $(1-\alpha)\lambda$ non-duplicated arrivals, and
the entirety of the $\alpha\lambda$ duplicated arrivals, are informative
(because two packets are produced but exactly one is informative),
\begin{equation}
	\lambda_{\mathrm{inf}} = \lambda\bigl[(1-\alpha)P_{\mathrm{inf},1}+\alpha\bigr],
	\qquad \mathbb{E}[T] = {1}/{\lambda_{\mathrm{inf}}}.
	\label{eq:ET-dup}
\end{equation}

The transform argument above is, however, insufficient to obtain the cross
moment $\mathbb{E}[ST\mid 1]$ and the second moment $\mathbb{E}[T^2\mid 1]$,
because these require the \emph{joint} survival function of
$({C}_1,{C}_2)$, but ${C}_1$ and ${C}_2$ are
not independent: with probability $\alpha$, the very packet that generates
$X_1$ is itself duplicated and therefore also generates $X_2$ at the same
instant (branch~$A$), while with the complementary probability $1-\alpha$ the
first duplicated arrival after $X_1$ occurs an independent delay
$L\sim\mathrm{Exp}(\lambda')$ later (branch~$B$), so $X_2=X_1+L$. This
coupling has no effect on the marginal law of ${C}_2$ used above,
but it does affect the joint law required for the moments, since $X_1$ and
$X_2$ can coincide.

Let $t:=s+w$ denote the elapsed time since the tagged departure, and recall
$X_1\sim\mathrm{Exp}(\lambda)$ from the candidate construction above; we
write $x_1$ for its realization to avoid confusion with the branch label
$A$. Conditional on the threatening arrival to queue~2 having occurred at
$X_2=y<t$, the survival probability
$F(y):=\Pr({C}_2>t\mid X_2=y)$ depends on the pre-existing state of
queue~2 at the reference epoch, which -- being uncorrelated with the
non-duplicated tagged packet -- is queue~2's own stationary state: busy,
with probability $\rho_2$, or idle, with probability $1-\rho_2$. If busy, let
$V$ denote the residual time, from the reference epoch, until queue~2's
pre-existing content clears; $V$ is exponential of rate $\nu_2$ by the same
memoryless sojourn-time property used for $R_j$ in~\eqref{eq:Rj}. Then
\begin{equation}
\begin{split}
    F(y) = \underbrace{\rho_2\,e^{-\nu_2 t}}_{\Pr(V\geq t)}
	+ \underbrace{\bigl[1-\rho_2 e^{-\nu_2 y}\bigr]e^{-\mu_2(t-y)}}_{\Pr(V\leq y)\,\Pr(Y_2>t-y)} \\
	+ \int_y^t \rho_2\nu_2\, e^{-\nu_2 v}\,e^{-\mu_2(t-v)}\,dv, \qquad y<t,
	\label{eq:Qy-int}
\end{split}	
\end{equation}
where the first term is the probability that the pre-existing content is
still present at $t$; the second is the probability that it has already
cleared by $y$, times the probability that a fresh service started at $y$
survives to $t$; and the integral accounts for the pre-existing content
clearing at some intermediate instant $v\in(y,t)$, after which a new service
started at $v$ must itself survive to $t$. Evaluating the integral and
collecting terms (using $\rho_2\nu_2=\lambda'-\rho_2\lambda'$) collapses
\eqref{eq:Qy-int} to
\begin{equation}
	F(y) = e^{-\nu_2 t} + e^{-\mu_2(t-y)} - e^{-\mu_2 t+\lambda' y}, \quad y<t,
	\label{eq:Qy}
\end{equation}
with $F(y)=1$ for $y\geq t$. Under branch~$A$, $X_2=x_1$ and the queue-2
survival factor is $F(x_1)$ directly; under branch~$B$, $X_2=x_1+L$ and the
factor must be averaged over $L$, which, splitting the integration domain at
$L=t-x_1$ (the point at which $x_1+L$ crosses $t$), gives
\begin{equation}
	\varphi_B(x_1) = \int_{x_1}^{t}\lambda' e^{-\lambda'(y-x_1)}F(y)\,dy + e^{-\lambda'(t-x_1)}, \quad x_1<t,
	\label{eq:phiB-branch}
\end{equation}
and $\varphi_B(x_1)=1$ for $x_1\geq t$. Weighting queue~1's own-queue survival
function, identical in form to $R_i(t\mid s)$ in~\eqref{eq:Ri}, since
queue~1's dynamics are themselves unaffected by duplication, by $F(x_1)$ on
branch~$A$ and by $\varphi_B(x_1)$ on branch~$B$, and integrating over the
arrival epoch $x_1$ with the same busy/empty split at $x_1=s$ used
in~\eqref{eq:Ri}, gives
\begin{IEEEeqnarray}{rCl}
P_A(w,s)
 &=& \int_0^s \lambda e^{-\lambda x_1}e^{-\mu_1 w}F(x_1)\,dx_1 \label{eq:PA}
\\
 &+& \int_s^t \lambda e^{-\lambda x_1}e^{-\mu_1(t-x_1)}F(x_1)\,dx_1
        + e^{-\lambda t}.
\nonumber\\
P_B(w,s)
 &=& \int_0^s \lambda e^{-\lambda x_1}e^{-\mu_1 w}\varphi_B(x_1)\,dx_1 \label{eq:PB}
\\
 &+& \int_s^t \lambda e^{-\lambda x_1}e^{-\mu_1(t-x_1)}
 \varphi_B(x_1)\,dx_1 + e^{-\lambda t},
\nonumber
\end{IEEEeqnarray}
and
\vspace{-0.4cm}
\begin{equation}
\begin{split}
    	P(w,s) = \alpha P_A(w,s) + (1-\alpha)P_B(w,s) \\
	= \Pr({C}_1>s+w,\ {C}_2>s+w\mid S=s),
	\label{eq:Pws}
\end{split}
\end{equation}
the joint survival function of the two candidate departures conditional on
the tagged system time. Setting $w=0$, integrating $P(0,s)$ against the
exponential density of $S$, and specializing reproduces
exactly~\eqref{eq:Pinf1-xf} after simplification, confirming the reduction
claimed above. All integrals in~\eqref{eq:Qy-int}--\eqref{eq:Pws} are, after
expansion, finite sums of terms $c\,x^ke^{rx}$ with $k\in\{0,1\}$, integrable
in closed form via $\int_a^b x^ke^{rx}dx$ and the stability-guaranteed
vanishing of the boundary term at infinity.

By the law of total expectation applied to $P(w,s)$,
\begin{IEEEeqnarray}{l}
    \mathbb{E}[ST\mid 1] = \frac{1}{P_{\mathrm{inf},1}}
	\int_0^\infty \!\!\!\nu_1e^{-\nu_1s}\,s\Bigl[\int_0^\infty \!\!P(w,s)\,dw\Bigr]ds, \qquad
    \label{eq:EST-case1}
	\\
	\mathbb{E}[T^2\mid 1] = \frac{2}{P_{\mathrm{inf},1}}
	\int_0^\infty \!\!\!\nu_1e^{-\nu_1s}\Bigl[\int_0^\infty \!\!w\,P(w,s)\,dw\Bigr]ds, \qquad
	\label{eq:ET2-case1}
\end{IEEEeqnarray}
and, since $P(w,s)$ is by construction a finite sum of terms $c\,w^ke^{-aw}$
with coefficients that are themselves finite sums of terms $c's^le^{-bs}$,
both nested integrals in~\eqref{eq:EST-case1} and~\eqref{eq:ET2-case1} reduce, after expansion, to
finite sums of the elementary form $p!/\alpha^{p+1}\cdot q!/\beta^{q+1}$,
exact and free of numerical quadrature.

\subsection*{Cases 2 and 3: duplicated record}

As both queues are FCFS, no packet
generated after a tagged duplicate can depart from \emph{either} queue before
the tagged duplicate's earliest copy does: any later-generated packet routed
through queue~1 necessarily departs after the tagged copy already present in
queue~1, and, if it is itself duplicated, its queue-2 copy likewise departs
after the tagged copy already present in queue~2. Of these duplicates, exactly one will be informative and the other will not, so no equivalent of
$P_{\mathrm{inf},1}$ is required for Cases~2 and~3; the relevant quantity is
instead the probability that queue~1's copy departs before queue~2's,
$\Pr(S_1<S_2)$.

Notably, since queue~2's arrivals are a sub-sample of queue~1's, although $S_1$ and $S_2$ are marginally exponentially distributed, they are also positively correlated, which rules out an elementary
independent-exponential race identity and requires the exact joint law of
the two system times. Let $n_1,n_2$ denote the number of packets present in
queue~1 and queue~2, respectively, at the instant the tagged packet is
duplicated. Their joint stationary distribution $\pi(n_1,n_2)$ comes from a bivariate Markov chain with transition rates $\lambda(1-\alpha)$ for a
non-duplicated arrival (queue~1 only), $\lambda\alpha$ for a simultaneous
arrival at both queues, and $\mu_1,\mu_2$ for the respective departures. 
Conditional on $(n_1,n_2)$, PASTA
gives $S_1\sim\mathrm{Erlang}(n_1{+}1,\mu_1)$ and
$S_2\sim\mathrm{Erlang}(n_2{+}1,\mu_2)$, the time to complete $n_1{+}1$,
respectively $n_2{+}1$, exponential services, so that
\begin{IEEEeqnarray}{rCl}
	\label{eq:PS1S2}
    	\Pr(S_1<S_2) &=& \sum_{n_1,n_2}\pi(n_1,n_2)\!\!\!\sum_{j=n_1+1}^{n_1+n_2+1}
	\binom{n_1{+}n_2{+}1}{j} \qquad \\
    &{\cdot}& \ p^j(1-p)^{n_1+n_2+1-j}, \quad p:=\frac{\mu_1}{\mu_1+\mu_2}, \nonumber
\end{IEEEeqnarray}
the inner sum being the exact win probability of an Erlang race, obtained by
conditioning on the order statistics of $n_1{+}n_2{+}2$ i.i.d.\ exponential
clocks assigned to queue~1 with probability $p$ and to queue~2 with
probability $1-p$.

With $s:=S_1$, $u:=S_2-S_1$ in Case~2 (respectively $s:=S_2$, $u:=S_1-S_2$ in
Case~3), the candidate survival function
$\Pr({C}_1>s+w,{C}_2>s+w\mid s,u)$ is built by the same
branch-$A$/branch-$B$ argument as~\eqref{eq:PA}--\eqref{eq:PB}, now applied
to the arrival $x_1$ that generates the post-duplication candidate, and
further split at $w=u$, the instant at which the losing copy departs. For
$w\leq u$ the losing queue is by definition still occupied by its own
pending copy, so within this interval only a departure from the reference
queue's own service process can possibly precede $s+w$: the condition on
the competing queue is trivially satisfied and the joint survival collapses
to the survival function of the reference queue alone. When queue~1 is the
reference (Case~2) this is simply the elementary single-queue survival
$R_1(w\mid s)$ of~\eqref{eq:Ri}; when queue~2 is the reference (Case~3), however, queue~2's own arrivals are already the thinned duplicate sub-stream, so its ``own'' survival is not $R_2(w\mid s)$ but a branch-$A$/branch-$B$ composite of the same kind as $F(\cdot)$ in~\eqref{eq:Qy}, restricted to $w\le u$.

For $w>u$, integrating over $x_1$ against the branch-$A$ factor $F(x_1)$
(queue~2's threat, now evaluated relative to the reference $s$) and
queue~1's own factor, split at the two thresholds $s$ and $s+u$ at which
each queue's pre-existing content clears, produces two exponential terms,
\begin{equation}
	P_A^{\mathrm{case}}(w\mid s,u) = C_1(s,u)\,e^{-(\mu_1+\mu_2)w} + C_2(s,u)\,e^{-\lambda w},
	\label{eq:PAgt}
\end{equation}
with $C_1,C_2$ collecting the contributions of the three regions "both
queues still holding their pre-existing content," "the reference queue has
cleared, the other has not," and "both cleared." Because branch~$A$ involves
a single shared arrival racing against two independent services, it is
symmetric under exchange of the reference and competing queues, and
\eqref{eq:PAgt} applies to both Case~2 (reference queue~1) and Case~3
(reference queue~2) with $\mu_1,\mu_2$ exchanged accordingly.

The branch-$B$ construction, in contrast, is \emph{not} symmetric between
the two cases. Along the arrival axis $x_1$, exactly two thresholds are
crossed before certainty sets in at $s+w$: the instant at which queue~1's
own pre-existing content clears, and the instant at which queue~2's own
pre-existing content clears. Which of the two coincides with $s$ and which
with $s+u$ depends on the case, $s+u=S_1$ in Case~2, $s+u=S_2$ in Case~3,
but a second, independent asymmetry governs which of the two per-queue
survival factors is elementary and which is itself a branch-$A$/branch-$B$
composite. Queue~1 receives the full, undivided Poisson stream of rate
$\lambda$, so its own busy/empty transition, wherever its threshold falls,
is the elementary factor already used in~\eqref{eq:Ri}. Queue~2, by
contrast, receives only the duplicated sub-stream, tied to queue~1's
arrivals through the same branch-$A$/branch-$B$ mechanism used to construct
$F(\cdot)$ in Case~1; consequently, \emph{whichever role queue~2 plays},
its survival factor is always a composite of $F(c):=
e^{-\mu_2c}$ and $J(c):=\frac{\mu_2}{\mu_2-\lambda'}e^{-\lambda'c}
-\frac{\lambda'}{\mu_2-\lambda'}e^{-\mu_2c}$, and it is this factor's own
threshold that requires extra bookkeeping.

In Case~2, queue~2 is the \emph{competing} queue, with threshold $s+u$;
its composite factor,
\begin{equation}
\psi_2(x_1) = F(w{-}u) + e^{-\lambda'(s{+}u{-}x_1)}\bigl[J(w{-}u)-F(w{-}u)\bigr],
\label{eq:psi2case2}
\end{equation}
with $x_1<s+u$ and $\psi_2(x_1)=J(s+w-x_1)$ for $x_1\geq s+u$, therefore remains
\emph{unchanged} across both the interval $x_1\in[0,s)$, where queue~1
(the reference, elementary factor) still holds its own pending copy, and
$x_1\in[s,s+u)$, where queue~1 has already cleared: both intervals lie
below queue~2's threshold $s+u$, so \eqref{eq:psi2case2} applies to both and the only transition visible between these two regions is in
queue~1's elementary factor, from $e^{-\mu_1w}$ to $e^{-\mu_1(s+w-x_1)}$.
The composite factor only switches once, from \eqref{eq:psi2case2} to
$J(s+w-x_1)$, at the boundary between the second and third region,
\begin{IEEEeqnarray}{l}
P_B^{\mathrm{case2}}(w\mid s,u) =
\int_0^s \lambda e^{-\lambda x_1}\,e^{-\mu_1w}\,\psi_2(x_1)\,dx_1 \label{eq:PBcase2} \\
\;\; +\!\int_s^{s+u} \!\!\lambda e^{-\lambda x_1}\,e^{-\mu_1(s+w-x_1)}\,\psi_2(x_1)\,dx_1 \nonumber \\
\;\;+\!\int_{s+u}^{s+w} \!\!\lambda e^{-\lambda x_1}\,e^{-\mu_1(s+w-x_1)}\,J(s{+}w{-}x_1)\,dx_1 
+ e^{-\lambda(s+w)}. \nonumber
\end{IEEEeqnarray}

In Case~3 the roles are exchanged: queue~2 is now the \emph{reference}
queue, with threshold $s$, so it is its composite factor, not queue~1's
elementary one, that switches between the first and second region, while
queue~1, now competing with threshold $s+u$, contributes the elementary
factor $e^{-\mu_1(w-u)}$ throughout $x_1<s+u$ and switches only once, to
$e^{-\mu_1(s+w-x_1)}$, at $x_1=s+u$. Mirroring~\eqref{eq:psi2case2}, queue~2's
composite factor is
\begin{equation}
\varphi_2^{<}(x_1) = F(w) + e^{-\lambda'(s{-}x_1)}\bigl[J(w){-}F(w)\bigr], \quad x_1{<}s, \;\;
\label{eq:phi2case3}
\end{equation}
and $\varphi_2^{\geq}(x_1)=J(s+w-x_1)$ for $x_1\geq s$, the same terminal
form as in Case~2, since once queue~2's own pending copy has cleared, its
subsequent behavior no longer depends on which case is being computed.
Combining the two elementary/composite factors across the three regions
gives the mirror image of~\eqref{eq:PBcase2},
\begin{IEEEeqnarray}{l}\label{eq:PBcase3}
P_B^{\mathrm{case3}}(w\mid s,u) =
\int_0^s \lambda e^{-\lambda x_1}\,e^{-\mu_1(w-u)}\,\varphi_2^{<}(x_1)\,dx_1 \\
\qquad +\int_s^{s+u} \lambda e^{-\lambda x_1}\,e^{-\mu_1(w-u)}\,\varphi_2^{\geq}(x_1)\,dx_1 \nonumber \\
\qquad +\int_{s+u}^{s+w} \lambda e^{-\lambda x_1}\,e^{-\mu_1(s+w-x_1)}\,\varphi_2^{\geq}(x_1)\,dx_1 
+ e^{-\lambda(s+w)}. \nonumber
\end{IEEEeqnarray}
$\varphi_2^{<}(x_1)$ and $\varphi_2^{\geq}(x_1)$ are, respectively, the
reference queue's own survival given that the competing arrival at $x_1$
found it still holding its pending copy ($x_1<s$) or already cleared
($x_1\geq s$) -- the Case-3 counterpart of $F(\cdot)$ in~\eqref{eq:Qy}, with
the roles of the reference and competing queue exchanged. In the first
integral queue~1 (competing) is still within its own pre-existing window
($x_1<s\le s+u$); the second integral covers the same window for queue~1
but with queue~2 (reference) already cleared, $s\leq x_1<s+u$; the third
covers $x_1\in[s+u,s+w)$, where queue~1 has also re-armed; the fourth
covers the event that no competing arrival occurs before $s+w$ at all. As
in~\eqref{eq:Qy-int}, every term in~\eqref{eq:PBcase3} is, after expansion,
a finite sum of monomials $c\,x_1^ke^{rx_1}$ integrable in closed form.

In short: the composite, branch-derived factor always belongs to queue~2,
by virtue of its arrivals being the thinned sub-stream; which pair of
regions it distinguishes, $(1,2)$ or $(2,3)$, depends only on whether
queue~2 is acting as the reference or the competing queue in the case at
hand.

Combining,
$P^{\mathrm{case}}(w\mid s,u)=\alpha P_A^{\mathrm{case}}+(1-\alpha)P_B^{\mathrm{case}}$
for $w>u$ (with the case-specific $P_A^{\mathrm{case}}$, $P_B^{\mathrm{case}}$
above for Case~3, and the analogous single-level construction for Case~2),
together with the single-queue survival for $w\leq u$, is in both cases a
finite sum of terms $c\,e^{-aw-bu}$ with coefficients depending on $s$.
Decomposing $\mathbb{E}[ST\mid s,u]$ and $\mathbb{E}[T^2\mid s,u]$,
obtained from $P^{\mathrm{case}}(w\mid s,u)$ by the same survival-to-moment
identities as~\eqref{eq:EST-case1} and~\eqref{eq:ET2-case1}, into monomials $C\,s^pu^qe^{-as-bu}$,
and integrating each monomial in closed form against the exact
$\mathrm{Erlang}(n_1{+}1,\mu_1)\times\mathrm{Erlang}(n_2{+}1,\mu_2)$ density
via the binomial expansion of $(s+u)^n$ together with the elementary Gamma
identity $\int_0^\infty x^me^{-cx}dx=m!/c^{m+1}$, yields
$\mathbb{E}[ST\mid n_1,n_2]$ and $\mathbb{E}[T^2\mid n_1,n_2]$ in analytical
form for every $(n_1,n_2)$; summing over $(n_1,n_2)$ weighted by
$\pi(n_1,n_2)$, a deterministic, geometrically convergent series, truncated
at negligible error, gives the exact case-conditional moments
$\mathbb{E}[ST\mid 2],\mathbb{E}[T^2\mid 2],\mathbb{E}[ST\mid 3],\mathbb{E}[T^2\mid 3]$,
without recourse to the independent-exponential averaging over $u$ and $s$
that the splitting model would otherwise permit.

\begin{figure}[tbp]
    \centering
    \includegraphics[width=\figw]{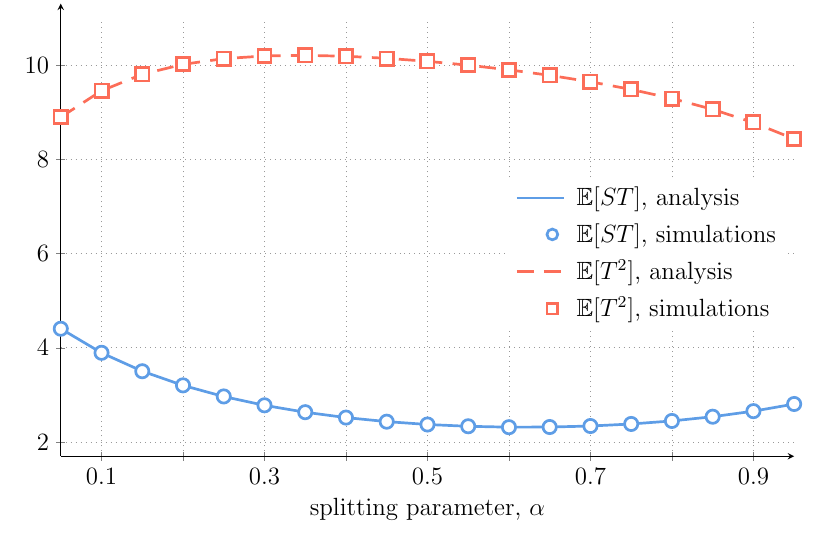}
    \caption{Cross-moment $\mathbb{E}[ST]$ and second moment $\mathbb{E}[T^2]$ for the
    split-merge model as a function of the splitting parameter $\alpha$:
    closed-form analytical result vs.\ Monte Carlo simulation, for
    $\lambda{=}0.5$, $\mu_1{=}1$, $\mu_2{=}0.8$.}
    \label{fig:est_vs_alpha_split}
\end{figure}

\begin{figure}[tbp]
    \centering
    \includegraphics[width=\figw]{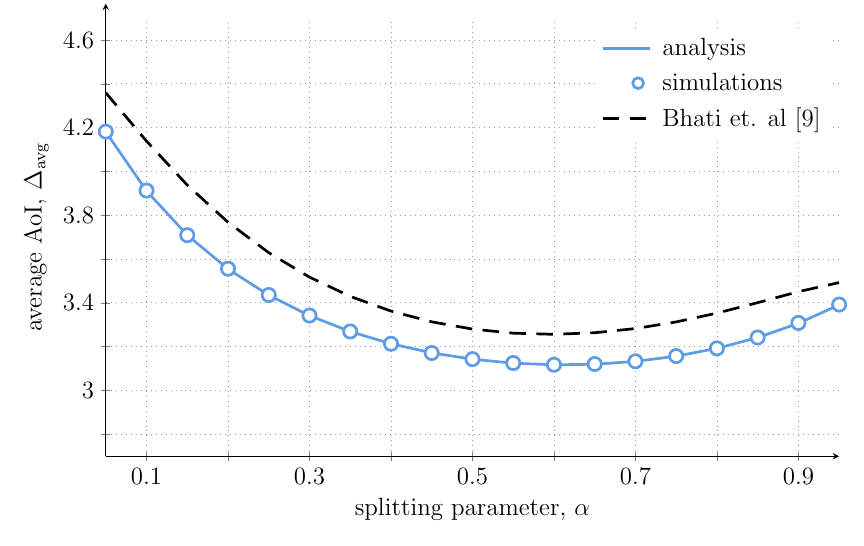}
    \caption{Average AoI $\Delta_{\mathrm{avg}}$ for the
    split-merge model as a function of the splitting parameter $\alpha$:
    closed-form analytical result vs.\ Monte Carlo simulation, for
    $\lambda{=}0.5$, $\mu_1{=}1$, $\mu_2{=}0.8$, also compared against Bhati~\textit{et al.}~\cite{bhati2021age}.}
    \label{fig:aoi_vs_alpha_split}
\end{figure}

The three cases are recombined through the general average-AoI formula of~\eqref{eq:aoi-general}, with weights given by the long-run share of
informative departures each case contributes,
\begin{IEEEeqnarray}{ll}
    	p_1 = \frac{\lambda(1-\alpha)P_{\mathrm{inf},1}}{\lambda_{\mathrm{inf}}}, \quad &
	p_2 = \frac{\lambda\alpha\,\Pr(S_1<S_2)}{\lambda_{\mathrm{inf}}}, 	\label{eq:weights-dup}
\\
	p_3 = \frac{\lambda\alpha\,\Pr(S_2<S_1)}{\lambda_{\mathrm{inf}}}, & \qquad \textrm{so that:} \nonumber
\end{IEEEeqnarray}
\begin{equation}
\begin{split}
    	\mathbb{E}[ST] = p_1\mathbb{E}[ST\mid 1]+p_2\mathbb{E}[ST\mid 2]+p_3\mathbb{E}[ST\mid 3],
	\\
	\mathbb{E}[T^2] = p_1\mathbb{E}[T^2\mid 1]+p_2\mathbb{E}[T^2\mid 2]+p_3\mathbb{E}[T^2\mid 3],
	\label{eq:final-moments-dup}
\end{split}
\end{equation}
which, together with~\eqref{eq:ET-dup}, provide all the ingredients required
to evaluate the average AoI of \eqref{eq:aoi-general} in analytical
form.

\begin{figure}[tbp]
    \centering
    \includegraphics[width=\figw]{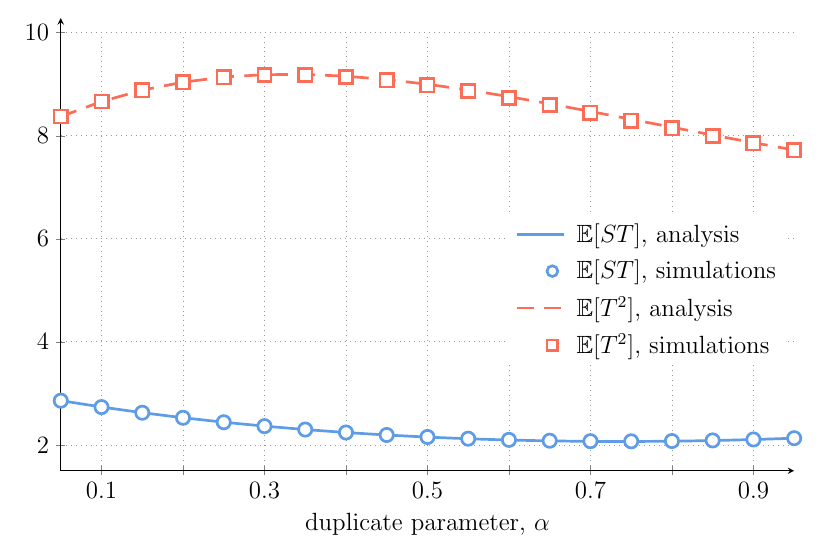}
    \vspace{-0.2cm}
    \caption{Cross-moment $\mathbb{E}[ST]$ and second moment $\mathbb{E}[T^2]$ for the
    duplicate-merge model as a function of the duplication parameter $\alpha$:
    closed-form analytical result vs.\ Monte Carlo simulation, for
    $\lambda{=}0.5$, $\mu_1{=}1$, $\mu_2{=}0.8$.}
    \label{fig:est_vs_alpha_dup}
\end{figure}

\begin{figure}[tbp]
    \centering
    \includegraphics[width=\figw]{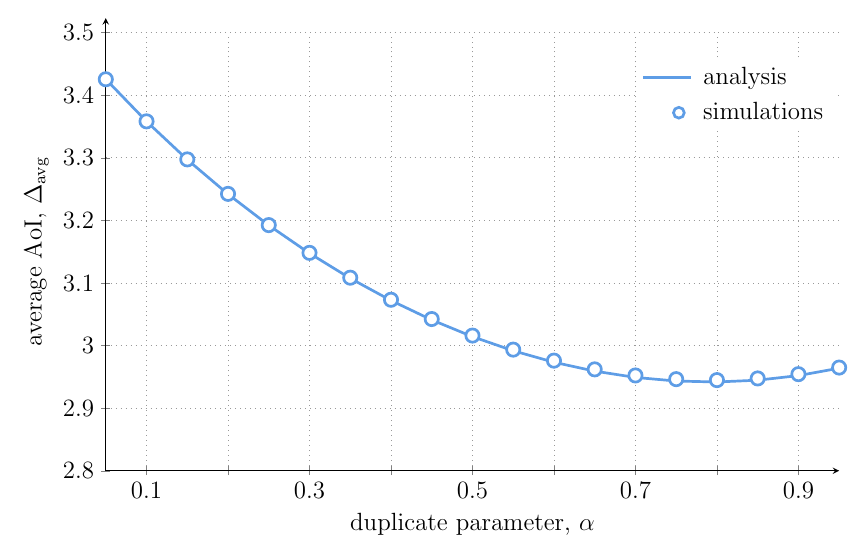}
    \caption{Average Age of Information $\Delta_{\mathrm{avg}}$ for the
    duplicate-merge model as a function of the duplication parameter $\alpha$:
    closed-form analytical result versus Monte Carlo simulation, for
    $\lambda=0.5$, $\mu_1=1$, $\mu_2=0.8$.}
    \label{fig:aoi_vs_alpha_dup}
\end{figure}

\section{Numerical Results}
\label{sec:result}

We first validate the closed-form moments of Sections~\ref{sec:split-model} and~\ref{sec:analysis_duplicate} and the resulting average AoI of~\eqref{eq:aoi-general}, against Monte Carlo simulation, using $\lambda=0.5$, $\mu_1=1$, $\mu_2=0.8$. For the split-merge model, Fig.~\ref{fig:est_vs_alpha_split} shows that $\mathbb{E}[ST]$ and $\mathbb{E}[T^2]$ match simulation exactly across the whole range of $\alpha$, confirming~\eqref{eq:EST-final} and~\eqref{eq:ET2}. Fig.~\ref{fig:aoi_vs_alpha_split} shows the resulting $\Delta_{\mathrm{avg}}$, again matching simulation; since the split-merge model with two heterogeneous M/M/1 queues under random routing reduces to the system of Bhati~\textit{et al.}~\cite{bhati2021age}, we also report their derivation for comparison, showing how it achieves close but not fully exact  results; in particular, they correctly reflect the optimal split rate $\alpha$ but not the AoI value that is slightly biased. For the duplicate-merge model, Figs.~\ref{fig:est_vs_alpha_dup} and~\ref{fig:aoi_vs_alpha_dup} show the same agreement with simulation, validating the case-decomposition of~\eqref{eq:weights-dup}--\eqref{eq:final-moments-dup} and, in particular, the joint law of $(S_1,S_2)$ used in~\eqref{eq:PS1S2} to weight Cases~2 and~3.

\begin{figure}[tbp]
    \centering
    \includegraphics[width=\figw]{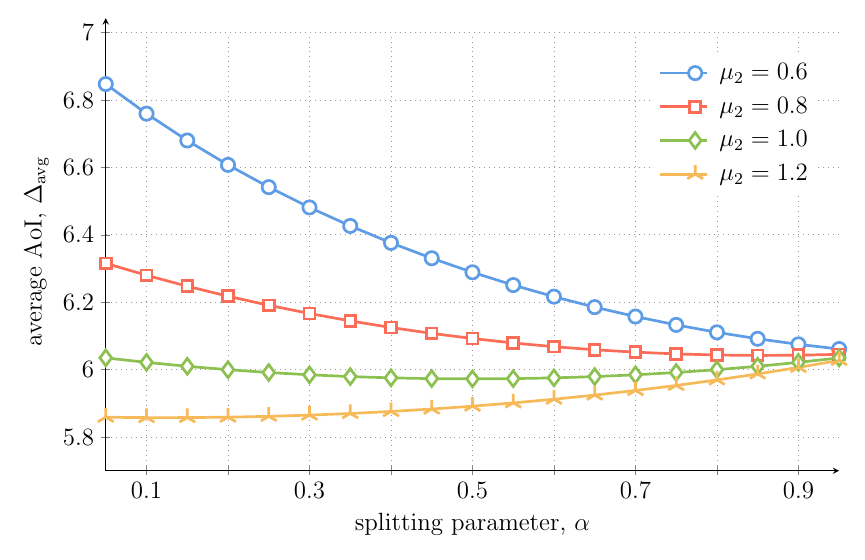}
    \caption{Average AoI $\Delta_{\mathrm{avg}}$ for the
    split-merge model as a function of the splitting parameter $\alpha$, for
    several values of $\mu_2$, at low arrival rate ($\lambda=0.2$, $\mu_1=1$).}
    \label{fig:lambda_low_split}
\end{figure}

\begin{figure}[tbp]
    \centering
    \includegraphics[width=\figw]{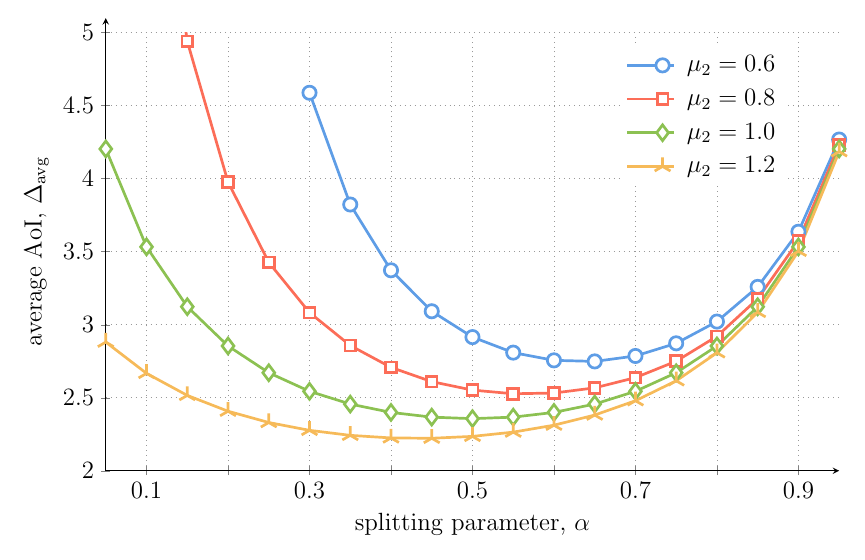}
    \caption{Average AoI $\Delta_{\mathrm{avg}}$ for the
    split-merge model as a function of the splitting parameter $\alpha$, for
    several values of $\mu_2$, at high arrival rate ($\lambda=0.8$, $\mu_1=1$).
    Curves for small $\mu_2$ are truncated where queue~2 becomes unstable.}
    \label{fig:lambda_high_split}
\end{figure}

\begin{figure}[tbp]
    \centering
    \includegraphics[width=\figw]{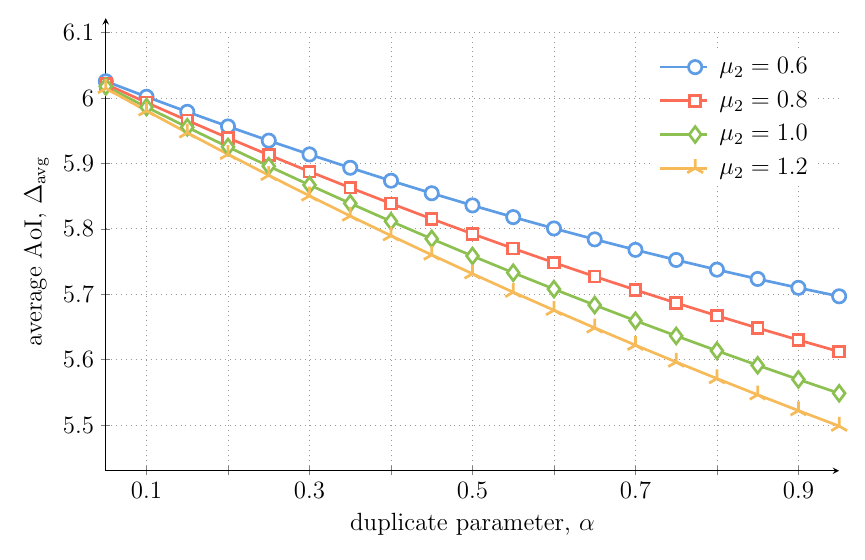}
    \caption{Average AoI $\Delta_{\mathrm{avg}}$ for the
    duplicate-merge model as a function of the duplication parameter $\alpha$,
    for several values of $\mu_2$, at low arrival rate ($\lambda=0.2$, $\mu_1=1$).}
    \label{fig:lambda_low_dup}
\end{figure}

\begin{figure}[tbp]
    \centering
    \includegraphics[width=\figw]{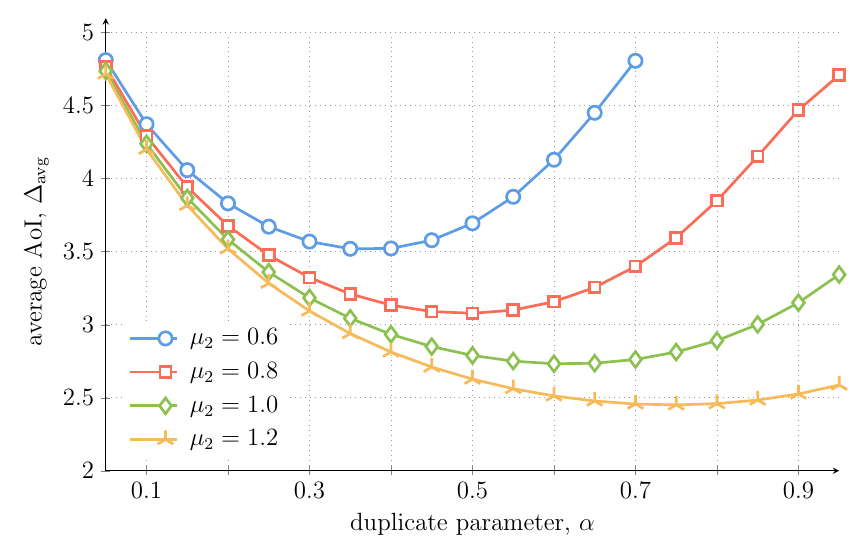}
    \caption{Average AoI $\Delta_{\mathrm{avg}}$ for the
    duplicate-merge model as a function of the duplication parameter $\alpha$,
    for several values of $\mu_2$, at high arrival rate ($\lambda=0.8$, $\mu_1=1$).
    Curves for small $\mu_2$ are truncated where queue~2 becomes unstable.}
    \label{fig:lambda_high_dup}
\end{figure}

Moreover, we explore how $\Delta_{\mathrm{avg}}$ depends on $\alpha$ and $\mu_2$ jointly, for a low and a high arrival rate ($\lambda=0.2$ and $\lambda=0.8$), with $\mu_1=1$ fixed. Fig.~\ref{fig:lambda_low_split} and Fig.~\ref{fig:lambda_high_split} show $\Delta_{\mathrm{avg}}$ versus $\alpha$ for the split-merge model, with a curve for each $\mu_2\in\{0.6,0.8,1.0,1.2\}$. At low $\lambda$ the curves are nearly flat and closely spaced, since the system is far from saturation and $\alpha$ has limited impact regardless of $\mu_2$. At high $\lambda$, each curve instead spans only the range of $\alpha$ for which both queues are stable, with small-$\mu_2$ curves truncated at small $\alpha$ (routing too little traffic away from queue~2 drives it toward instability); away from this boundary, the curves develop a pronounced interior minimum. Fig.~\ref{fig:lambda_low_dup} and Fig.~\ref{fig:lambda_high_dup} show the analogous behavior for the duplicate-merge model, where instead queue~1 is always stable ($\lambda<\mu_1$) while queue~2, with rate $\alpha\lambda$, becomes unstable for $\alpha$ too \emph{large}: at high $\lambda$ the small-$\mu_2$ curves are thus truncated at large $\alpha$, the opposite side from split-merge. At low $\lambda$ this constraint is rarely binding, and larger $\alpha$ is generally beneficial, since duplication then adds redundancy without seriously threatening queue~2's stability.

Table~\ref{tab:lambda_aoi_opt} and Table~\ref{tab:lambda_aoi_opt_duplicate} report $\lambda^*$ and the corresponding minimum AoI as functions of $\alpha$ and $\mu_2$, for the split-merge and duplicate-merge model respectively ($\mu_1=1$ in both cases). For split-merge, $\lambda^*$ (and $\Delta_{\mathrm{avg}}^*$) peaks at an interior $\alpha$ that shifts with $\mu_2$, from $\alpha\approx0.7$ at $\mu_2=0.4$ down to exactly $\alpha=0.5$ at $\mu_2=\mu_1$, reflecting the optimal load balance between the two servers. For duplicate-merge, $\lambda^*$ varies over a narrower range ($0.44$--$0.69$, versus $0.25$--$1.07$ for splitting), but its optimal $\alpha$ moves in the \emph{opposite} direction, from $\alpha\approx0.3$ at $\mu_2=0.4$ to $\alpha\approx0.7$ at $\mu_2=1.0$: a faster secondary server makes heavier duplication worthwhile. In both models, the achievable minimum $\Delta_{\mathrm{avg}}^*$ decreases monotonically with $\mu_2$, reflecting the diminishing bottleneck of the secondary server.

\begin{figure}[t!]
    \centering
    \includegraphics[width=\figw]{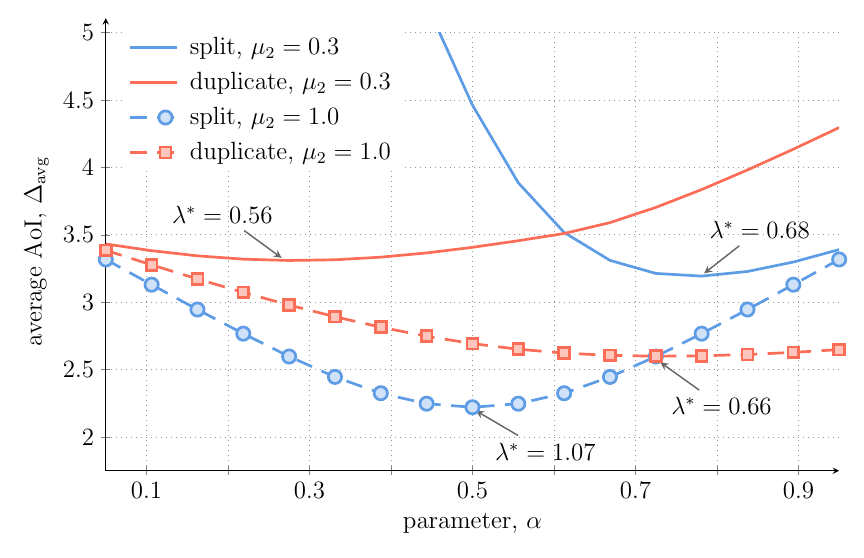}
    \caption{Optimal average AoI as a function of $\alpha$ for the split and duplicate routing policies, for $\mu_1=1.0$, and $\mu_2\in \{0.3,1.0 \}$. 
    }
    \label{fig:confronto}
\end{figure}

Finally, Fig.~\ref{fig:confronto} compares the two policies directly, plotting $\Delta_{\mathrm{avg}}^*=\Delta_{\mathrm{avg}}(\lambda^*)$ versus $\alpha$ for two representative values of $\mu_2$. Split-merge achieves the lowest attainable AoI overall, but by a margin that depends strongly on $\mu_2$. When the secondary server is slow ($\mu_2=0.3$), the two policies are nearly equivalent at their optima ($3.194$ at $\alpha=0.78$ for split-merge versus $3.310$ at $\alpha=0.27$ for duplicate-merge), so the choice of policy matters little, though $\alpha$ still needs careful tuning since the split-merge curve diverges rapidly away from its optimum. When the secondary server is fast ($\mu_2=1.0$), split-merge is substantially better ($2.221$ at $\alpha=0.50$ versus $2.600$ at $\alpha=0.72$), since splitting fully exploits the extra capacity while duplication wastes it serving the same packet twice; this optimum, $\alpha=0.50$ with $\lambda^*=1.067$, corresponds to $0.533$ per queue, which is consistent with what is found in~\cite{bhati2021age}, despite the approximation introduced there.

\begin{table}[tbp]
\centering
\caption{Optimal arrival rate $\lambda^*$ and corresponding minimum AoI, for $\mu_1=1$ (split-merge model).}
\label{tab:lambda_aoi_opt}
\begin{tabular}{cc@{\hspace{7pt}}cc@{\hspace{7pt}}cc@{\hspace{7pt}}cc@{\hspace{7pt}}c}
\toprule
\multirow{2}{*}{$\alpha$} & \multicolumn{2}{c}{$\mu_2=0.4$} & \multicolumn{2}{c}{$\mu_2=0.6$} & \multicolumn{2}{c}{$\mu_2=0.8$} & \multicolumn{2}{c}{$\mu_2=1.0$} \\
\cmidrule(lr){2-3} \cmidrule(lr){4-5} \cmidrule(lr){6-7} \cmidrule(lr){8-9}
 & $\lambda^*$ & $\Delta_{\rm avg}^*$ & $\lambda^*$ &  $\Delta_{\rm avg}^*$ & $\lambda^*$ &  $\Delta_{\rm avg}^*$ & $\lambda^*$ &  $\Delta_{\rm avg}^*$ \\
\midrule
0.1 & 0.249 & 7.687 & 0.372 & 5.172 & 0.494 & 3.911 & 0.615 & 3.151 \\
0.3 & 0.382 & 5.614 & 0.556 & 3.909 & 0.719 & 3.049 & 0.873 & 2.527 \\
0.5 & 0.691 & 3.701 & 0.835 & 2.911 & 0.956 & 2.490 & 1.067 & 2.221 \\
0.7 & 0.752 & 3.043 & 0.805 & 2.786 & 0.845 & 2.631 & 0.873 & 2.527 \\
0.9 & 0.605 & 3.268 & 0.610 & 3.213 & 0.613 & 3.176 & 0.615 & 3.151 \\
\bottomrule
\end{tabular}
\end{table}

\begin{table}[tbp]
\centering
\caption{Optimal arrival rate $\lambda^*$ and corresponding minimum AoI, for $\mu_1=1$ (duplicate-merge model).}
\label{tab:lambda_aoi_opt_duplicate}
\begin{tabular}{cc@{\hspace{7pt}}cc@{\hspace{7pt}}cc@{\hspace{7pt}}cc@{\hspace{7pt}}c}
\toprule
\multirow{2}{*}{$\alpha$} & \multicolumn{2}{c}{$\mu_2=0.4$} & \multicolumn{2}{c}{$\mu_2=0.6$} & \multicolumn{2}{c}{$\mu_2=0.8$} & \multicolumn{2}{c}{$\mu_2=1.0$} \\
\cmidrule(lr){2-3} \cmidrule(lr){4-5} \cmidrule(lr){6-7} \cmidrule(lr){8-9}
 & $\lambda^*$ & $\Delta_{\rm avg}^*$ & $\lambda^*$ &  $\Delta_{\rm avg}^*$ & $\lambda^*$ &  $\Delta_{\rm avg}^*$ & $\lambda^*$ &  $\Delta_{\rm avg}^*$ \\
\midrule
0.1 & 0.553 & 3.364 & 0.559 & 3.330 & 0.563 & 3.306 & 0.565 & 3.289 \\
0.3 & 0.572 & 3.225 & 0.601 & 3.096 & 0.622 & 3.005 & 0.637 & 2.940 \\
0.5 & 0.545 & 3.254 & 0.594 & 3.017 & 0.642 & 2.834 & 0.685 & 2.693 \\
0.7 & 0.505 & 3.395 & 0.554 & 3.077 & 0.609 & 2.817 & 0.669 & 2.601 \\
0.9 & 0.444 & 3.577 & 0.512 & 3.207 & 0.561 & 2.898 & 0.619 & 2.630 \\
\bottomrule
\end{tabular}
\end{table}

\section{Conclusion}
\label{sec:conclusion}

We introduced an analytical framework to characterize AoI in multi-path queueing topologies with converging server streams. By investigating split-merge and duplicate-merge M/M/1 network architectures, we resolved critical analytical limitations in prior studies. Specifically, our exact closed-form solutions corrected inaccuracies found in existing split-merge models, while our novel derivation for the duplicate-merge paradigm established the first formal analysis of this scenario.  

Our findings provide practical guidelines for tuning randomized routing probabilities and traffic injection rates to minimize average AoI. These insights offer robust architectural foundations for ultra-reliable low-latency communications, edge computing, and real-time cyber-physical systems.

\section*{Use of AI disclosure}
All analytical derivations, proofs, and simulation code were developed by the authors. ChatGPT was used solely as an independent implementation aid to generate an \emph{additional} Monte Carlo simulator for cross-validation.

\IEEEtriggeratref{12}
\bibliographystyle{IEEEtran}
\bibliography{references}

@ARTICLE{yates2021age,
  author={Yates, Roy D. and Sun, Yin and Brown, D. Richard and Kaul, Sanjit K. and Modiano, Eytan and Ulukus, Sennur},
  journal={IEEE J. Sel. Areas Commun.}, 
  title={Age of Information: An Introduction and Survey}, 
  year={2021},
  volume={39},
  number={5},
  pages={1183-1210}}

@article{kam2015effect,
  title={Effect of message transmission path diversity on status age},
  author={Kam, Clement and Kompella, Sastry and Nguyen, Gam D and Ephremides, Anthony},
  journal={IEEE Trans.\ Inf.\ Theory},
  volume={62},
  number={3},
  pages={1360--1374},
  year={2015},
}

@article{chiariotti2022age,
  title={Age of information in multihop connections with tributary traffic and no preemption},
  author={Chiariotti, Federico and Vikhrova, Olga and Soret, Beatriz and Popovski, Petar},
  journal={IEEE Trans. Commun.},
  volume={70},
  number={10},
  pages={6718--6733},
  year={2022},
}

@article{fidler2023statistical,
  title={Statistical age-of-information bounds for parallel systems: When do independent channels make a difference?},
  author={Fidler, Markus and Champati, Jaya Prakash and Widmer, Joerg and Noroozi, Mahsa},
  journal={IEEE J. Sel. Areas Inf. Theory},
  volume={4},
  pages={591--606},
  year={2023},
  publisher={IEEE}
}

@inproceedings{bhati2021age,
  title={On the age of information of a queuing system with heterogeneous servers},
  author={Bhati, Anhad and Pillai, Sibi Raj B and Vaze, Rahul},
  booktitle={Proc. Nat. Conf. Commun. (NCC)},
  pages={1--6},
  year={2021},
}

@article{moltafet2020age,
  title={On the Age of Information in Multi-Source Queueing Models},
  author={Moltafet, Mohammad and Leinonen, Markus and Codreanu, Marian},
  journal={IEEE Trans. Commun.},
  volume={68},
  number={8},
  pages={5003--5017},
  year={2020},
}

@article{moltafet2021average,
  title={Average {AoI} in multi-source systems with source-aware packet management},
  author={Moltafet, Mohammad and Leinonen, Markus and Codreanu, Marian},
  journal={IEEE Trans. Commun.},
  volume={69},
  number={2},
  pages={1121--1133},
  year={2021},
}

@inproceedings{kaul2012real,
  title={Real-time status: How often should one update?},
  author={Kaul, Sanjit and Yates, Roy and Gruteser, Marco},
  booktitle={Proc. IEEE Conf. Comp. Commun. (INFOCOM)},
  pages={2731--2735},
  year={2012},
}

@article{lefebvre2026optimal,
  title={Optimal control of a queueing system},
  author={Lefebvre, Mario and Yaghoubi, Roozbeh},
  journal={Optimization},
  volume={75},
  number={2},
  pages={351--364},
  year={2026},
  publisher={Taylor \& Francis}
}

@article{reddy2025age,
  title={Age of information analysis for queueing systems with proactive obsolete packet management: multi-hop, multi-source, and priority mechanisms},
  author={Reddy, Y Arun Kumar and Venkatesh, TG},
  journal={IEEE Trans. Netw. Sci. Eng.},
  year={2025},
  volume = 12, number = 5, 
pages = {4316--4328},
}

@article{chen2024timeliness,
  title={Timeliness of status update system: The effect of parallel transmission using heterogeneous updating devices},
  author={Chen, Zhengchuan and Lang, Kang and Pappas, Nikolaos and Yang, Howard H and Wang, Min and Tian, Zhong and Quek, Tony QS},
  journal={IEEE Trans. Commun.},
  volume={72},
  number={11},
  pages={7093--7107},
  year={2024},
}

@inproceedings{kaul2020timely,
  title={Timely updates by multiple sources: The {M/M/1} queue revisited},
  author={Kaul, Sanjit K and Yates, Roy D},
  booktitle={Proc. Ann. Conf. Inf. Sci. Syst. (CISS)},
  pages={1--6},
  year={2020},
}

@article{yates2018age,
  title={The age of information: Real-time status updating by multiple sources},
  author={Yates, Roy D and Kaul, Sanjit K},
  journal={IEEE Trans. Inf. Theory},
  volume={65},
  number={3},
  pages={1807--1827},
  year={2019},
}

@article{tong2022age,
  title={Age-of-information oriented scheduling for multichannel {IoT} systems with correlated sources},
  author={Tong, Jingwen and Fu, Liqun and Han, Zhu},
  journal={IEEE Trans. Wireless Commun.},
  volume={21},
  number={11},
  pages={9775--9790},
  year={2022},
}

@article{liang2025analysis,
  title={Analysis of age of information in {CSMA} network with correlated sources},
  author={Liang, Long and Zhou, Siyuan},
  journal={Electronics},
  volume={14},
  number={13},
  pages={2688},
  year={2025},
  publisher={MDPI}
}

@article{zancanaro2023modeling,
  title={Modeling value of information in remote sensing from correlated sources},
  author={Zancanaro, Alberto and Cisotto, Giulia and Badia, Leonardo},
  journal={Comp. Commun.},
  volume={203},
  pages={289--297},
  year={2023},
  publisher={Elsevier}
}

@article{fidler20252d,
  title={{2D-AoI}: Age-of-information of distributed sensors for spatio-temporal processes},
  author={Fidler, Markus and Gallistl, Flavio and Champati, Jaya Prakash and Widmer, Joerg},
  journal={IEEE Trans. Commun.},
  volume={74},
  pages={645--661},
  year={2025},
  publisher={IEEE}
}

@inproceedings{tripathi2022optimizing,
  title={Optimizing age of information with correlated sources},
  author={Tripathi, Vishrant and Modiano, Eytan},
  pages={41--50},
  booktitle = {Proc. ACM MobiHoc},
  year={2022}
}

@inproceedings{mancuso2022stateful,
  title={Stateful versus stateless selection of edge or cloud servers under latency constraints},
  author={Mancuso, Vincenzo and Castagno, Paolo and Sereno, Matteo and Marsan, Marco Ajmone},
  booktitle={Proc. IEEE Int. Symp. World Wireless Mobile Multimedia Netw. (WoWMoM)},
  pages={110--119},
  year={2022},
}

@article{cohen2023dynamic,
  title={Dynamic service provisioning in the edge-cloud continuum with bounded resources},
  author={Cohen, Itamar and Chiasserini, Carla Fabiana and Giaccone, Paolo and Scalosub, Gabriel},
  journal={IEEE/ACM Trans. Netw.},
  volume={31},
  number={6},
  pages={3096--3111},
  year={2023},
}

@article{tang2023age,
  title={Age of information of multi-user mobile-edge computing systems},
  author={Tang, Zhifeng and Sun, Zhuo and Yang, Nan and Zhou, Xiangyun},
  journal={IEEE Open J. Commun. Soc.},
  volume={4},
  pages={1600--1614},
  year={2023},
  publisher={IEEE}
}

@article{popovski2022perspective,
  title={A perspective on time toward wireless {6G}},
  author={Popovski, Petar and Chiariotti, Federico and Huang, Kaibin and Kal{\o}r, Anders E and Kountouris, Marios and Pappas, Nikolaos and Soret, Beatriz},
  journal={Proc. IEEE},
  volume={110},
  number={8},
  pages={1116--1146},
  year={2022},
}

@article{jin2022review,
  title={A review of intelligent computation offloading in multiaccess edge computing},
  author={Jin, Hengli and Gregory, Mark A and Li, Shuo},
  journal={IEEE Access},
  volume={10},
  pages={71481--71495},
  year={2022},
  publisher={IEEE}
}

@article{chakraborti2019higher,
  title={Higher-order moments using the survival function: The alternative expectation formula},
  author={Chakraborti, Subhabrata and Jardim, Felipe and Epprecht, Eugenio},
  journal={Am. Stat.},
  year={2019},
  publisher={Taylor \& Francis}
}

\end{document}